# Beyond the Betz Theory – Blockage, Wake Mixing and Turbulence

Takafumi Nishino[*]

*Department of Engineering Science, University of Oxford, OX1 3PJ, UK*

*Summary*: Recent analytical models concerning the limiting efficiency of marine hydrokinetic (MHK) turbines are reviewed with an emphasis on the significance of blockages (of local as well as global flow passages) and wake mixing. Also discussed is the efficiency of power generation from fully developed turbulent open channel flows. These issues are primarily concerned with the design/optimization of tidal turbine arrays; however, some of them are relevant to wind turbines as well.

## Introduction

One of the key issues in tidal (and ocean-current) power generation is to properly understand the efficiency; not only the efficiency of each device but also the efficiency of device arrays or farms as a whole. A traditional way to estimate the limit of power generation from fluid flow is to use the so-called "Betz theory" (according to [1] the theory seems to have been developed independently by Lanchester, Betz and Joukowsky in the 1910's and 1920's). To better understand the efficiency of hydrokinetic turbines in practical situations, however, we need to consider several important factors that are not considered in the original Betz theory.

## Blockage and Wake Mixing Effects

The so-called "channel blockage" effect [2, 3] is one of the most influential factors to the limiting efficiency of hydrokinetic turbines when their ambient flow passage is confined in some form; for example, when turbines are installed in a relatively shallow water channel. By considering the conservation of mass, momentum and energy not only in the flow passing through the turbine cross-section (i.e. core flow) but also in the flow not passing through it (i.e. bypass flow), it can be shown that the limit of power generation from confined flow is proportional to $(1-B)^{-2}$, where $B$ is the ratio of turbine- to channel-cross-sectional areas.

The above model, which can be seen as a confined flow version of the Betz theory, yields a good estimation of the limiting efficiency of not only a single device but also a cross-stream array (or fence) of devices when they are regularly arrayed across an entire channel cross-section. When devices are arrayed only across a part of the cross-section, however, we need to think about (at least) two different types of blockages, namely the local and global blockages, $B_L$ and $B_G$ [4]. For example, if we consider a large number of devices arrayed only across a part of an infinitely wide channel (hence $B_G = 0$ but $B_L \neq 0$) and assume that all flow events around each device take place much faster than the horizontal expansion of flow around the entire array, it can be shown [4] that the limit of power generation may increase from the Betz limit of 59.3% (of the kinetic energy of undisturbed incoming flow) up to another limit of 79.8% depending on $B_L$. A more recent study [5] has shown that this "partial fence" model can be further extended, or generalized, by better accounting for the interaction of device- and array-scale flow events (so that the model can predict the efficiency of short as well as long fences).

Another influential factor to the efficiency of hydrokinetic turbines is wake mixing. It is widely known that wake mixing often plays a key role in deciding the efficiency of multiple fences, where the performance of downstream fences can be significantly affected by the wake of upstream fences (unless the streamwise gaps between them are large enough). A recent study [6] has shown, however, that wake mixing may also affect the limiting efficiency of a single fence and even a single device. This can be theoretically shown by considering energy transfer (due to mixing) between the bypass and core flows in the near-wake region and the attendant heat loss. This "near-wake mixing" effect might be negligible for a single device as its near-wake region is usually limited to only a few device-diameters (depending on the blockage). For a cross-stream array of devices, however, this effect is essential since the near-wake region of the entire array becomes much longer (compared to the scale of each device) as the number of devices in the array increases [5].

## Farm Efficiency in Turbulent Open Channel Flows

While all theories/models mentioned above are concerned with turbines subject to a uniform inviscid inflow, practical hydrokinetic turbines are usually placed in highly turbulent shear flows. To understand what kinds of turbines/farms are really "efficient" in such practical environments, we need to consider the balance between the

---

[*] Corresponding author.
*Email address*: takafumi.nishino@eng.ox.ac.uk

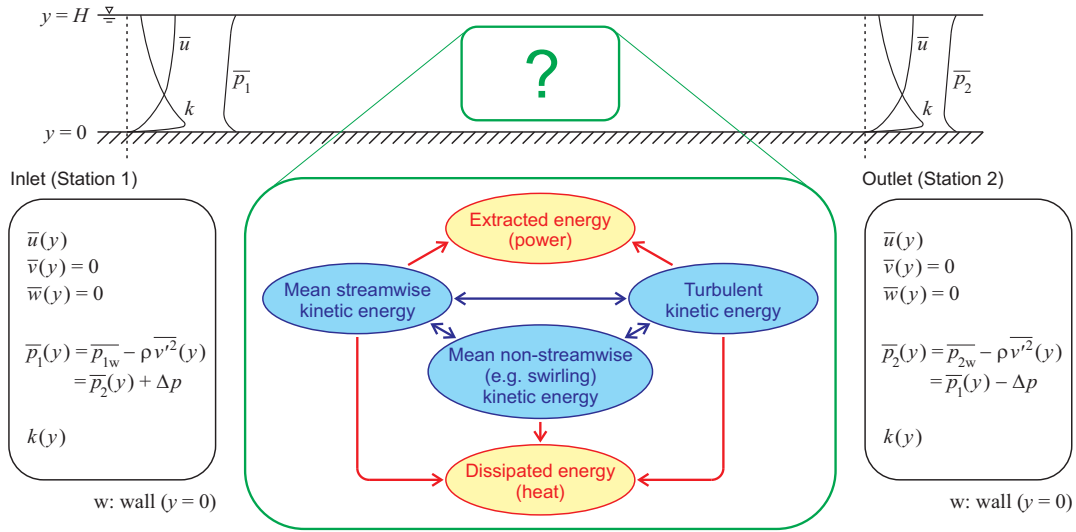

Fig. 1. Hydrokinetic power generation from a fully developed turbulent open channel flow.

energy extracted from the flow (as useful power) and that dissipated (into heat). Figure 1 shows a conceptual diagram of hydrokinetic power generation from a fully developed turbulent open channel flow (for simplicity here we employ the "rigid-lid" assumption and ignore any long-time-scale unsteadiness, e.g. due to tides). If we consider installing hydrokinetic turbines/farms in the middle of a long channel and assume that the flow returns to its fully-developed state before it reaches the outlet of the channel, the (exact) balances of mean-flow kinetic energy (including both streamwise and non-streamwise components) and turbulent kinetic energy in the region between the inlet and outlet of the channel can be described as

(mean-flow kinetic energy: MKE) $\quad P_{TKE}^{MKE} + P_{heat}^{MKE} + P_{power}^{MKE} = \langle \overline{u} \rangle HW \Delta p$

(turbulent kinetic energy: TKE) $\quad -P_{TKE}^{MKE} + P_{heat}^{TKE} + P_{power}^{TKE} = 0$

where $P_{TKE}^{MKE}$ represents the (net) energy transfer from MKE to TKE, $P_{heat}^{MKE}$ and $P_{heat}^{TKE}$ the dissipated energy from MKE and TKE, $P_{power}^{MKE}$ and $P_{power}^{TKE}$ the extracted energy from MKE and TKE, $\Delta p$ the pressure drop from the inlet to the outlet, $\langle \overline{u} \rangle$ the velocity averaged across the water height $H$, and $W$ ($\gg H$) is the channel width.

It should be noted that the earlier theoretical models [2-6] consider the extraction of energy only from MKE, whereas in reality some energy could be extracted from TKE as well. Also of importance is that the installation of devices/farms should change the values of not only $P_{power}^{MKE}$ and $P_{power}^{TKE}$ (from zero to non-zero values) but also $P_{TKE}^{MKE}$, $P_{heat}^{MKE}$ and $P_{heat}^{TKE}$, depending on how the devices/farms alter the flow around them. From the viewpoint of energy loss in turbulent channel flow, "efficient" arrays of devices or farms might be designed as a sort of flow control system that reduces the level of (near-wall) turbulence and thus the dissipation of energy in the flow.

*Acknowledgements:*
The author acknowledges the support of the Oxford Martin School, University of Oxford.